\documentclass[conference,letterpaper]{IEEEtran}

\usepackage[utf8]{inputenc} 
\usepackage[T1]{fontenc} 
\usepackage{url} 
\usepackage{ifthen} 
\usepackage{cite} 
\usepackage[cmex10]{amsmath} 

\usepackage[cmex10]{amsmath}
\usepackage{amssymb}
\usepackage{xcolor}
\usepackage{graphicx}

\usepackage{array}
\usepackage{enumitem} 
\usepackage{verbatim} 
\usepackage{bbold} 

\usepackage{bbm} 
\usepackage{mathtools} 

\newtheorem{thm}{\bf Theorem}

\newtheorem{lem}[thm]{\bf Lemma}

\newtheorem{rem}{\bf Remark}

\newenvironment{proof}{\noindent{\bf Proof:}}{$\hfill\circ$\\[0\baselineskip]}

\newcommand{\E}{\mathbb{E}}

\newcommand{\beq}{\begin{equation}}
\newcommand{\eeq}{\end{equation}}

\newcommand{\ind}{\mathbbm{1}}

\DeclareMathOperator*{\argmax}{arg\,max}

\makeatletter

\newcommand{\Rmnum}[1]{\expandafter\@slowromancap\romannumeral #1@}
\makeatother

\DeclareMathOperator*{\pr}{\mathbbm{P}}  

\newcommand{\bsA}{{\boldsymbol A}}

\newcommand{\bsS}{{\boldsymbol S}}
\newcommand{\bsX}{{\boldsymbol X}}

\newcommand{\caS}{{\mathcal S}}
\newcommand{\caN}{{\mathcal N}}

\newcommand{\caT}{{\mathcal T}}
\newcommand{\caW}{{\mathcal W}}
\newcommand{\caQ}{{\mathcal Q}}
\newcommand{\caV}{{\mathcal V}}
\newcommand{\mv}{\textsf{MV}}
\newcommand{\wmv}{\textsf{WMV}}
\newcommand{\sign}{\text{sign}}

\begin{document}
\title{Crowdsourced Labeling for
\\ Worker-Task Specialization Model}


\author{%
  \IEEEauthorblockN{Doyeon Kim and Hye Won Chung}
  \IEEEauthorblockA{School of Electrical Engineering\\
                    KAIST\\
                    Email: \{highlowzz, hwchung\}@kaist.ac.kr}
}


\maketitle

\begin{abstract}
We consider crowdsourced labeling under a $d$-type worker-task specialization model, where each worker and task is associated with one particular type among a finite set of types and a worker provides a more reliable answer to tasks of the matched type than to tasks of unmatched types. We design an inference algorithm that recovers binary task labels (up to any given recovery accuracy) by using worker clustering, worker skill estimation and weighted majority voting. The designed inference algorithm does not require any information about worker/task types, and achieves any targeted recovery accuracy with the best known performance (minimum number of queries per task). 
\footnote{This work was supported in part by National
Research Foundation of Korea under Grant 2017R1E1A1A01076340; in part
by the Ministry of Science and ICT, South Korea, under the ITRC support
program under Grant IITP-2021-2018-0-01402; and in part by the Institute of
Information and Communications Technology Planning \& Evaluation (IITP)
grant funded by the Korea Government MSIT under Grant 2020-0-00626.
}
\end{abstract}


%




\IEEEpeerreviewmaketitle

\section{Introduction}
We consider the problem of crowdsourced labeling, which has diverse applications in image labeling, video annotation, and character recognition~\cite{raykar2010learning,von2008recaptcha,welinder2010multidimensional}.
Workers in the crowdsourcing system are given simple tasks and asked to provide a binary label to each assigned task. 
Since workers may provide incorrect labels to some of the tasks and worker reliabilities are usually unknown, the main challenge in the crowdsourced labeling is to infer true labels from noisy answers collected from workers of unknown reliabilities. 

To resolve such challenges and to design inference algorithms with provable performance guarantees, many previous works considered a simple yet meaningful error model for workers' answers. One of the most widely studied model is the single-coin Dawid-Skene model~\cite{dawid1979maximum}, where each worker is modeled by his/her own reliability level and the worker provides a correct answer to any task with probability depending on the worker's reliability level, regardless of the types of assigned tasks. For such a model, various inference algorithms were proposed to first estimate the worker reliabilities from the collected answers and to use them to infer correct labels by using expectation maximization (EM)\cite{gao2013minimax,liu2012variational,zhou2012learning}, message passing \cite{karger2014budget}, or spectral method \cite{dalvi2013aggregating, zhang2014spectral}. 
However, this error model does not capture some realistic scenarios where worker's ability to provide a correct label could change depending on the types of the assigned tasks and the workers' expertise \cite{8437703,9174227,kim2020crowdsourced,shah2020permutation}. 

In this work, we consider a $d$-type specialization model, which was introduced in~\cite{shah2018reducing}. This model assumes that each worker and each task is associated with a single type (among $d$ different types), and a worker provides an answer better than a random guess if the task type matches the worker type and otherwise, the worker just provides a random guess. 
The inference algorithm proposed in~\cite{shah2018reducing} is composed of two stages. At the first stage, the workers are clustered based on similarity on their answers, and at the second stage the task label is estimated by first finding a cluster of the matched type and aggregating the answers only from the chosen cluster while ignoring the answers from other clusters. 

In this work, we generalize the $d$-type specialization model to the case where a worker provides an answer better than a random guess with probability $q\in[1/2,p)$ even when the worker type and the task type does not match. When the types are matched, the answer is correct with higher probability $p\in(q,1]$. Different from the algorithm in~\cite{shah2018reducing}, we do not throw away the answers from the cluster of unmatched type but use the answers with proper weights to achieve the optimal accuracy in the label estimation. We propose two algorithms in this paper. Our first algorithm does not require any information on the worker/task types but the parameters $(p,q)$, and it achieves the best known performance, regardless of the regimes of the reliability parameters $(p,q)$ or the number of types $d$. We then propose our second algorithm which does not require even $(p,q)$ values. In this algorithm, the parameters are estimated from the workers' answers and used to estimate the correct labels. We empirically show that our second algorithm achieve as good performance as the first algorithm in diverse parameter regimes. 
Furthermore, we empirically demonstrate that under the generalized $d$-type specialization model our two proposed algorithms outperform the state-of-the-art inference algorithms developed for the Dawid-Skene model. 

\section{Problem Formulation}\label{sec2}
In this work, we consider a $d$-type specialization model for crowdsourced labeling. 
We assume that there exists $m$ binary tasks and $n$ workers. Denote the set of tasks and the set of workers by $\caT$ and $\caW$, respectively. Let $\caW_z$ denote the set of workers of type $z\in[d]$. For $i \in \caT$, let $a_i \in \lbrace -1, 1 \rbrace$ denote the true label of the $i$-th task, and let $t_i, w_j \in [d]$ denote the type of the $i$-th task and that of the $j$-th worker, respectively, where $[d]:= \{ 1, \dots, d \}$.
We assume that the type of each task and the type of each worker are uniformly distributed over $[d]$.
The set of workers assigned to task $i$ is denoted by $\caN_i$. Let $m_{ij}$ be the $j$-th worker's answer to the task $i$. If task $i$ is not assigned to worker $j$, then $m_{ij}=0$, and if it is assigned 
\beq \label{eq2.1}
{m_{ij}} = 
\begin{cases}
a_i & \text{with probability $f_{ij}$}, \\
-a_i & \text{with probability $1-f_{ij}$}.
\end{cases}
\eeq
We assume that $m_{ij}$'s are independent for all $i,j$. The $d$-type specialization model we consider further assumes that
\beq
f_{ij}=
\begin{cases}
p,&\text{if }t_i=w_j, \\
q,&\text{o.w.}
\end{cases}
\eeq
where $p> q\geq 1/2$.
Different from~\cite{shah2018reducing} where the value $q$ was fixed to 1/2, here we consider a general $q\in[1/2,p)$.

For $i \in \caT$, let $\hat a_i \in \{-1, 1\}$ denote the inferred label of the $i$-th task.
The performance metric we consider is the expected fraction of errors in the inferred labels, i.e., $\E[\frac{1}{m}\sum_{i=1}^m\mathbbm{1}(\hat{a}_i\neq a_i)]=\frac{1}{m}\sum_{i=1}^m\pr(\hat{a}_i\neq a_i)$.
We aim to minimize the number of queries per task, achieving
\beq\label{eqn:err}
\frac{1}{m}\sum_{i=1}^m\pr(\hat{a}_i\neq a_i)\leq\alpha_c,\quad\text{for some }\alpha_c\in(0,1).
\eeq

\section{Performance Baselines}\label{sec3}
In this section, we first review performance baselines of previous works and outline our contributions.
\subsection{Oracle Weighted Majority Voting and Majority Voting}
As the first performance baseline, we consider a general weighted majority voting, which aggregates answers with weights to generate the label estimate. 
For weighted majority voting, the decision is given by
\beq\label{eqn:wmv_estimator}
\hat{a}_i^{\wmv}=\sign\left(\sum_{j\in \caN_i}\mu_{ij}m_{ij} \right),
\eeq
where  $\mu_{ij}$ is the weight for the answer from the $j$-th worker to the $i$-th task.
By using Hoeffding's inequality (or Corollary 5 in~\cite{li2014error}), it can be shown that the weighted majority voting guarantees
\beq\label{eqn:wmv_P}
\pr(\hat{a}_i^{\wmv}\neq a_i)\leq \exp\left(-\frac{\gamma^2_{\wmv}}{2}|\caN_i|\right)
\eeq
where 
\beq\label{eqn:wmv_P1}
\gamma_{\wmv}=\frac{\sum_{j\in\caN_i} \mu_{ij}(2f_{ij}-1)}{\|\mu_{i*}\|_2 \cdot\sqrt{|\caN_i|}}
\eeq
 for $\mu_{i*} = ( \mu_{i1}, \dots,\mu_{i \vert \caN_i \vert} )$.
By Cauchy-Schwarz inequality, the weight $\mu_{ij}$ that maximizes $\gamma_\wmv$ is $\mu_{ij}\propto (2f_{ij}-1)$.

When we choose $\caN_i \subset \caW$ at random, effectively, $1/d$ fraction of answers are given with fidelity $f_{ij}=p$ and the rest with $f_{ij}=q$.
Thus, when $\{f_{ij}\}$ is known, i.e., when the task types $\{t_i\}$ and the worker types $\{w_j\}$ as well as the reliability parameters $(p,q)$ are known at the inference algorithm, by choosing $\mu_{ij}\propto (2f_{ij}-1)$ the oracle weighted majority voting can achieve~\eqref{eqn:wmv_P} with
$
\gamma_{\wmv}^*=\frac{\sqrt{(2p-1)^2+(d-1)(2q-1)^2}}{\sqrt{d}}.
$
The required number of queries per task to achieve~\eqref{eqn:err} for the oracle weighted majority voting is thus
\beq\label{eqn:wmv_Ld}
L_{\sf oracle}=\frac{2d}{(2p-1)^2+(d-1)(2q-1)^2}\ln\left(\frac{1}{\alpha_c}\right).
\eeq

As another baseline, we can consider the simple majority voting that aggregates all the answers with equal weights, i.e.,
$
\hat{a}_i^{\mv}=\sign\left(\sum_{j\in \caN_i}m_{ij} \right).
$
The majority voting gives
$
\pr(\hat{a}_i^{\mv}\neq a_i)\leq \exp\left(-\frac{\gamma_{\mv}^2}{2}|\caN_i|\right)
$
where
$
\gamma_{\mv}=\frac{((2p-1)+(d-1)(2q-1))}{{d}}.
$
To achieve the targeted recovery accuracy~\eqref{eqn:err} with the majority voting, the required number of queries per task is
\beq\label{eqn:mv_Ld}
L_{\sf mv}=\frac{2d^2}{((2p-1)+(d-1)(2q-1))^2}\ln\left(\frac{1}{\alpha_c}\right).
\eeq
We can easily check the $L_{\sf oracle}$ in~\eqref{eqn:wmv_Ld} is less than or equal to $L_{\sf mv}$ in~\eqref{eqn:mv_Ld}.
However, the oracle result is achievable when the worker types and the task types as well as reliability parameters $(p,q)$ are all known to the inference algorithm. 

\subsection{Inference Algorithm from~\cite{shah2018reducing}: Clustering and Majority Voting from the Workers of a Matched Cluster}\label{sec:prev_3}
 We review the algorithm in~\cite{shah2018reducing}, proposed for the $d$-type specialization model with $p>q=1/2$.
The parameters $\zeta$, $r$, and $l$ of this algorithm can be chosen later to guarantee the recovery condition~\eqref{eqn:err}. 

\medskip 
\noindent\textbf{Algorithm~\cite{shah2018reducing}}: This algorithm is composed of two stages.
\begin{itemize} [leftmargin=*] 
\item\textit{Stage 1 (Clustering Workers by Types):} Let $\caS\subset \caT$ represent randomly chosen $r$ tasks from the set $\caT$. 
Assign each task in $\caS$ to all $n$ workers. Given the answers $m_{ij}$ for $i\in\caS$, cluster workers \textit{sequentially}: for a worker $j\in[n]$ if there exists a cluster of workers $\caQ\subset[j-1]$ such that for each $j'\in \caQ$
\beq\label{eqn:cluster1}
\frac{1}{r}\sum_{i\in\caS}\mathbbm{1}(m_{ij}=m_{ij'})>\zeta,
\eeq
then assign $j$ to $\caQ$; otherwise, create a new cluster containing $j$. Let $\{\caV_1,\dots,\caV_c\}$ be the resulting clusters of $[n]$ workers. 
For each task $i\in\caT\backslash \caS$ and cluster $z\in[c]$, assign task $i$ to $l$ workers sampled uniformly at random from the set $\caV_z$. The total number of workers assigned to task $i$ is $lc$.
\item\textit{Stage 2 (Type Matching and Majority Voting):} For each task $i\in\caT$, find a cluster of the matched type by
\beq\label{eqn:palg_typematching}
z^*(i)=\argmax_{z\in[c]}\left|\sum_{j\in\caN_i \cap \caV_z }m_{ij}\right|,
\eeq
and estimate the label for the task $i$ by the majority voting from the answers only from the set $\caV_{z^*(i)}$:
\beq\label{eqn:palg_est}
\hat{a}_i=\sign\left(\sum_{j\in \caN_i \cap \caV_{z^*(i)}} m_{ij}\right).
\eeq 
\end{itemize}
\medskip

The main idea of this algorithm is to cluster workers by finding subsets of workers having similarity (larger than some threshold $\zeta$) in their answers for the initially assigned $|\caS|=r$ tasks. After assigning the rest of the tasks $\caT\backslash \caS$ to total $lc$ workers from $c$ clusters, the final decision is made by the majority voting from the answers only from one cluster believed to be composed of workers having the same type as the task.
The parameters $\zeta$, $r$, and $l$ of this algorithm can be chosen to guarantee the recovery condition~\eqref{eqn:err}. We note that the choice of $\zeta$, which is $\frac{1}{2}+\frac{(p-q)^2}{d}$ in~\cite{shah2018reducing}, requires a prior knowledge of the model parameter $p,q$. 

We can easily generalize the analysis of this original algorithm to a general $q\geq 1/2$ by selecting a proper choice of $\zeta$, $r$, $n$ and $l$, and can show that the required number of queries per task $\frac{1}{m}(nr+ld(m-r))$ to achieve the recovery condition~\eqref{eqn:err} can be bounded as
\beq\label{eqn:Ld_palg}
L_{\sf type}= \min \left\{ \frac{2d}{\frac{(p-q)^2}{2}+\frac{(2q-1)^2}{2}} \ln\frac{6d+3}{\alpha_c}, \frac{2d}{\frac{(p-q)^2}{2}} \ln \frac{6d}{\alpha_c} \right\}
\eeq
when $r=\frac{d^2}{2(p-q)^4}\ln\frac{3n(n-1)}{2\alpha_c}$, $n\geq \max\left\{8d\ln\frac{3d}{\alpha_c},L_{\sf type}\right\}$,  $m\geq cn^3$, and  $l = \frac{1}{2d} L_{\sf type} $ for some constant $c>0$. 
\medskip
\begin{rem}[Our contributions]
When $q=1/2$ and $d$ is large, the clustering-based algorithm can guarantee the recovery condition~\eqref{eqn:err} with the number of queries per task scaling as $\frac{d}{(2p-1)^2}\ln \frac{d}{\alpha_c}$, whereas the majority voting requires $\frac{d^2}{(2p-1)^2}\ln\frac{1}{\alpha_c}$ queries per task. This demonstrates the benefit of using the clustering-based algorithm for $q=1/2$. 
The gain comes from aggregating a selected subset of answers from a matched cluster; in contrast, even though the majority voting aggregates almost $d$ times large number of answers, since $(d-1)l$ answers are just random guesses, these answers degrade the overall inference performance, especially when $d$ is large. 
On the other hand, for any $q>1/2$, the clustering-based algorithm requires much more number of queries $\frac{d}{(p-q)^2+(2q-1)^2}\ln \frac{d}{\alpha_c}$ compared to that of majority voting $\frac{d^2}{((2p-1)+(d-1)(2q-1))^2}\ln\frac{1}{\alpha_c}\approx\frac{1}{(2q-1)^2}\ln\frac{1}{\alpha_c}$, since the clustering-based algorithm does not utilize the $(d-1)l$ answers from unmatched clusters even though these answers can still provide some useful information about the true task label when $q>1/2$. 
Motivated by this observation, in the next section we propose two new algorithms, still based on clustering, but that aggregates the answers from all the clusters with proper weights. 
In particular, our second algorithm uses a new clustering method based on semidefinite programming (SDP) \cite{ames2014guaranteed,hajek2016achieving,vinayak2016similarity,lee2020hypergraph}, which does not require the knowledge of the reliability parameters $p,q$, and we also suggest estimators $\hat{p},\hat{q}$ calculated from the clustering result, which then can be used for weighted majority voting of workers' answers. 
\end{rem}
\section{Main Results}\label{sec4}
\subsection{First Algorithm: When Parameters $(p,q)$ are Known}
We first consider the case when $(p,q)$ are known so that we can use the optimal weighted majority voting after the clustering step in {\it Stage 1} of Algorithm~\cite{shah2018reducing}. 
With general $q\in[1/2,p)$, for the optimal weighted majority voting {\it Stage 2} of Algorithm~\cite{shah2018reducing} should be changed as below. 

\medskip
\noindent\textbf{Algorithm 1 (for the known $(p,q)$ case)}: This algorithm is composed of two stages. \textit{Stage 1} for worker clustering is the same as that of Algorithm \cite{shah2018reducing}, which is summarized in Section~\ref{sec:prev_3}. \textit{Stage 2} is modified as below.
\begin{itemize} [leftmargin=*]
\item\textit{Stage 2 (Type Matching and Weighted Majority Voting):} For each task $i\in\caT$, find a cluster of the matched type $z^*(i)$ by~\eqref{eqn:palg_typematching}
and set the weights $\mu_{ij}$ for answers $m_{ij}$, $j\in\caN_i$, by
\beq\label{eqn:weights_cluster}
\mu_{ij}=
\begin{cases}
2p-1,&\text{ for }j\in \caV_{z^*(i)},\\
2q-1,&\text{ for }j\in \caN_i \backslash \caV_{z^*(i)}.
\end{cases}
\eeq
Estimate the label for the task $i$ by the weighted majority voting~\eqref{eqn:wmv_estimator} with weights~\eqref{eqn:weights_cluster} based on the worker clustering and  the type matching. 
\end{itemize}

\begin{thm}
	\textit{With Algorithm 1, for any $\alpha_c \in (0,1)$, when $m \ge cn^3$ for some constant $c>0$, the recovery of task labels is guaranteed with the expected accuracy~\eqref{eqn:err}, with the number of queries per task
	\beq \label{eqn:Ld_alg1}
	L_{\sf Alg1} = \frac{2d}{{(p-q)^2}/{2} + \gamma_u} \ln \frac{6d+3}{\alpha_c}
	\eeq
	where
	\beq \label{eqn:gammau}
	\gamma_u=\frac{(2(2p-1)(2q-1)+(d-2)(2q-1)^2)^2}{2((2p-1)^2+(d-1)(2q-1)^2)}.
	\eeq}
\end{thm}
\begin{rem} Note that Algorithm 1 guarantees the recovery condition~\eqref{eqn:err} with a reduced number $L_{\sf Alg1}$ of queries per task compared to that of {Algorithm~\cite{shah2018reducing}} in~\eqref{eqn:Ld_palg}. Especially, the gap increases as $q(<p)$ increases.
Compared to the required number~\eqref{eqn:mv_Ld} of queries for majority voting, we can see that the proposed algorithm requires the same order $\Theta \left(\ln\frac{d}{\alpha_c} \right)$ (ignoring the $\ln d$ overhead) of queries when $q>1/2$ and $d\to\infty$, while that of {Algorithm~\cite{shah2018reducing}} required $\Theta \left(d\ln \frac{d}{\alpha_c}\right)$ queries per task. 
\end{rem}

\begin{proof}
With the two-stage algorithm, the workers are first clustered, and for a given task, the cluster of the matched type is inferred. 
We first analyze the clustering error. For any two workers $(a,b)$ having the same type, $\pr(m_{ia}=m_{ib}|w_a=w_b)=\frac{p^2+(1-p)^2}{d}+\frac{(d-1)(q^2+(1-q)^2)}{d},
$
while for two workers of different types, $\pr(m_{ia}=m_{ib}|w_a\neq w_b)=\frac{2(pq+(1-p)(1-q))}{d}+\frac{(d-2)(q^2+(1-q)^2)}{d}.
$
By setting $\zeta$ in~\eqref{eqn:cluster1} as the mean of the two values, we can bound $\pr(\frac{1}{r}\sum_{i\in\caS}\mathbbm{1}(m_{ia}=m_{ib})<\zeta| w_a=w_b)\leq \exp\left(-\frac{2(p-q)^4}{d^2}r\right)$ and $\pr(\frac{1}{r}\sum_{i\in\caS}\mathbbm{1}(m_{ia}=m_{ib})\geq \zeta| w_a\neq w_b)\leq \exp\left(-\frac{2(p-q)^4}{d^2}r\right)$ by using Chernoff bound. By union bound, the clustering error is then bounded by
\beq\label{eqn:cluster_er}
\pr(\text{Clustering error})\leq { n \choose 2} \exp\left(-\frac{2(p-q)^4}{d^2}r\right).
\eeq 
We also need to guarantee that the number of workers per type is at least $l$. Since the number of workers per type is distributed by $\text{Binomial}(n,\frac{1}{d})$, by using the Chernoff bound and the union bound,
\beq
\begin{split}\label{eqn:type_er}
\pr(\cup_{z\in [d]} \{|\mathcal{V}_z|\leq l\})
\leq d\exp\left(-\frac{1}{2}\left(1-\frac{ld}{n}\right)^2 \frac{n}{d}\right).
\end{split}
\eeq
Next, we bound the type matching error.
Let  $S_{iz}:=\sum_{j\in\mathcal{N}_i\cap \mathcal{W}_z}\mathbbm{1}(m_{ij}=+1)$. Note that $S_{iz}$ is distributed by Binomial($|\mathcal{N}_i\cap \mathcal{W}_z|,p$) if $t_i=z$ and $a_i=1$; Binomial($|\mathcal{N}_i\cap \mathcal{W}_z|,q$) if $t_i\neq z$ and $a_i=1$; Binomial($|\mathcal{N}_i\cap \mathcal{W}_z|,1-p$) if $t_i=z$ and $a_i=-1$;  and Binomial($|\mathcal{N}_i\cap \mathcal{W}_z|,1-q$) if $t_i\neq z$ and $a_i=-1$.
Therefore, if $S_{iz}$ is concentrated around its mean by $\frac{1}{2}(p-q)$, then~\eqref{eqn:palg_typematching} provides the correctly matched type. 
By the union bound over $z\in[d]$, the type matching error is thus bounded above by
\beq\label{eqn:type_match_err}
\pr(z^*(i)\neq t_i)\leq 2d\exp\left(-\frac{(p-q)^2l}{2}\right).
\eeq
We then analyze the label estimation error. When the clustering is perfect but the type matching is wrong, the weight defined in~\eqref{eqn:weights_cluster} is not equal to the desired weight $\mu_{ij} = 2 f_{ij} - 1$, and the estimation error is bounded above by the case when the weight is higher ($\mu_{ij}=2p-1)$ for a cluster that is incorrectly matched to the task, and lower $(\mu_{ij}=2q-1)$ for the cluster having the same type as the task, i.e., $\pr \left( \hat a_i^{\textsf{WMV}} \ne a_i \right) \le \exp \left( - \gamma_u l \right)$ where for $\gamma_u$ in~\eqref{eqn:gammau}. On the other hand, when the clustering and type matching is all correct, the estimation error for $\hat a_i^{\sf WMV}$ is equal to that of the oracle weighted majority voting, $\exp (-\gamma_m l)$ for $\gamma_m = \frac{(2p-1)^2+(d-1)(2q-1)^2}{2}$. It can be shown that $\exp(-\gamma_m l) \le \exp (-((p-q)^2/2 + \gamma_u) l )$.
By combining the above analysis, the expected fraction of label errors $\E\left[\frac{1}{m}\sum_{i=1}^m\mathbbm{1}(\hat{a}_i\neq a_i)\right]$ is bounded above by
\beq
\begin{split}
&{ n \choose 2} \exp\left(-\frac{2(p-q)^4}{d^2}r\right)+d\exp\left(-\frac{1}{2}\left(1-\frac{ld}{n}\right)^2 \frac{n}{d}\right)\\
&\qquad\quad+(2d+1)\exp\left(-\frac{(p-q)^2l}{2}\right) \cdot \exp(-\gamma_u l).
\end{split}
\eeq
To limit the fraction of errors to $\alpha_c$, we can choose $r=\frac{d^2}{2(p-q)^4}\ln\frac{3n(n-1)}{2\alpha_c}$, $l= \frac{1}{(p-q)^2/2+\gamma_u}\ln\frac{6d+3}{\alpha_c}$ and $n\geq \max\left\{8d\ln\frac{3d}{\alpha_c}, \frac{2d}{(p-q)^2/2+\gamma_u}\ln\frac{6d+3}{\alpha_c}\right\}$. The total number of queries per task is $\frac{1}{m}(nr+ld(m-r))\leq ld+\frac{nr}{m}$, and the second term is dominated by the first term when $m\ge c n^3$ for some constant $c>0$. Thus, the total number of queries per task is bounded by $L_{\sf Alg1}$ in~\eqref{eqn:Ld_alg1}.
\end{proof}
\vspace{-0.3cm}
\subsection{Second Algorithm: When Parameters $(p,q)$ are Unknown}
In this section, we propose a new algorithm that does not require the knowledge of reliability parameters $(p,q)$. 
For the purpose, we change both the clustering algorithm in \textit{Stage 1} and the weighted majority voting in \textit{Stage 2} of {Algorithm 1}.

\medskip
\noindent\textbf{Algorithm 2 (for the unknown $(p,q)$ case)}: 
\begin{itemize}[leftmargin=*]
\item\textit{Stage 1 (Clustering Workers by Types):} 
\begin{itemize}
\item Data preparation: after assigning each of $|\caS|=r$ tasks to all $n$ workers, construct a data matrix $\boldsymbol{S}\in\{-1,1\}^{r\times n}$, and define the similarity matrix $\bsA=\bsS^T\bsS$ while zeroing out the diagonal term of $\bsA$.
\item Parameter estimation for within-cluster and cross-cluster edge densities: compute and find the largest two eigenvalues of $\bsA$. Denote them by $\lambda_1$ and $\lambda_2$. Set $\hat{p}_c=\frac{\lambda_1+(d-1)\lambda_2}{n-d}$ and $\hat{q}_c=\frac{\lambda_1-\lambda_2}{n}$. 
\item Clustering Based on SDP (Algorithm 1 in~\cite{lee2020hypergraph}): With a tuning parameter $\lambda=\frac{\hat{p}_c+\hat{q}_c}{2}$, solve the SDP problem 
\beq
    \begin{split}
        \label{eq9}
        \max_{\mathbf{X} \in \mathbb{R}^{n \times n}} \ &\langle \mathbf{A} - \lambda \mathbf{1}_{n \times n}, \mathbf{X} \rangle \\
        \textnormal{subject to } &\mathbf{X} \succeq \mathbf{O};\ \langle \mathbf{I}_n, \mathbf{X} \rangle = n; \\
        &0 \leq \mathbf{X}_{ij} \leq 1,\ \forall i, j \in [n].
    \end{split}
\eeq
Employ the approximate $k$-medoids clustering algorithm (Algorithm 1 in~\cite{fei2018exponential}) on the optimal solution $\hat \bsX_{\sf SDP}$ of SDP to extract an explicit clustering, $\{\caV_1, \dots, \caV_d\}$.
\end{itemize}
\item \textit{Stage 2 (Type Matching and Weighted Majority Voting):}
for each task $i\in\caT$, find the cluster of matched type $z^*(i)$ by~\eqref{eqn:palg_typematching}.
\begin{itemize}
\item Randomly split each cluster: for each $z\in[d]$, randomly split the workers in $\caV_z$ into $\caV^{(1)}_z$ and $\caV^{(2)}_z$ with probability $\beta$ and $1-\beta$ respectively, where $\beta>0$ is a small enough probability. 
Let $\caW^{(1)}=\cup_{z=1}^d \caV^{(1)}_z$, $\caW^{(2)}=\cup_{z=1}^d \caV^{(2)}_z$, and $(\caV^{(1)}_z)^c=\caW^{(1)}\backslash \caV^{(1)}_z$ for $z\in[d]$.
\item Estimate ${p}$ and ${q}$: for $z^*(i)$ in~\eqref{eqn:palg_typematching}, define $\mathcal{M}(i):=\caN_i \cap \caV^{(1)}_{z^*(i)} $ and $\mathcal{U}(i):=\caN_i \cap (\caV^{(1)}_{z^*(i)})^c$, i.e, $\mathcal{M}(i)$ ($\mathcal{U}(i)$) is the set of workers in $\caW^{(1)}$ who answered for the task $i$ and are believed to have the matched (unmatched) type. 
Define $\hat{p}=\frac{1}{m}\sum_{i=1}^m\hat{p}_i$ and $\hat{q}=\frac{1}{m}\sum_{i=1}^m\hat{q}_i$ where 
\beq
\begin{split}\nonumber
\hat{p}_i&=\max\left\{\frac{\sum\limits_{j\in\mathcal{M}(i) }\ind(m_{ij}=1)}{|\mathcal{M}(i)|},\frac{\sum\limits_{j\in\mathcal{M}(i) }\ind(m_{ij}=-1)}{|\mathcal{M}(i) |}\right\},\\
\hat{q}_i&=\max\left\{\frac{\sum\limits_{j\in\mathcal{U}(i) }\ind(m_{ij}=1)}{|\mathcal{U}(i)|},\frac{\sum\limits_{j\in\mathcal{U}(i) }\ind(m_{ij}=-1)}{|\mathcal{U}(i) |}\right\}.
\end{split}
\eeq


\item Set the weights $\mu_{ij}$ as in~\eqref{eqn:weights_cluster} by replacing $p$ by $\hat{p}$ and $q$ by $\hat{q}$, and estimate the label for the task $i$ by the weighted majority voting
$
\hat{a}_i^{\wmv}=\sign\left(\sum_{j\in \caN_i \cap \caW^{(2)}}\mu_{ij}m_{ij} \right).
$
\end{itemize}
\end{itemize}
\begin{rem} 
We remark that {Algorithm 2} does not require any prior information about reliability parameters $(p,q)$ nor the task/worker types.
\textit{Stage 1} of {Algorithm 2} clusters workers by applying SDP to the similarity matrix with the tuning parameter $\lambda$ chosen from the data, and \textit{Stage 2} of {Algorithm 2} first finds a matched  cluster and uses this information to obtain the estimates $(\hat{p},\hat{q})$ of the model parameters $(p,q)$, which then can be used for the weighted majority voting. 
\end{rem}

The performance of the clustering algorithm is guaranteed by the lemma below.

\begin{lem} \label{lem1} \textit{
Suppose the tuning parameter $\lambda$ in the SDP~\eqref{eq9} obeys the bound $ \frac{1}{4} rp_m + \frac{3}{4} rp_u \leq \lambda \leq \frac{3}{4} rp_m + \frac{1}{4} rp_u$ where $p_m:=((2p-1)^2+(d-1)(2q-1)^2)/d$ and $p_u:=(2(2p-1)(2q-1)+(d-2)(2q-1)^2)/d$. Then, there is a universal constant $c_1>0$ such that Stage 1 of Algorithm 2 achieves the strong consistency with probability at least $1-4n^{-1}$ if $r \geq c_1 \frac{d^2 (\ln n)^2}{(p_m-p_u)^2}$.}
\end{lem}

\section{Numerical Results} \label{sec5}
\begin{figure}
	\centering
	\includegraphics[width=\columnwidth]{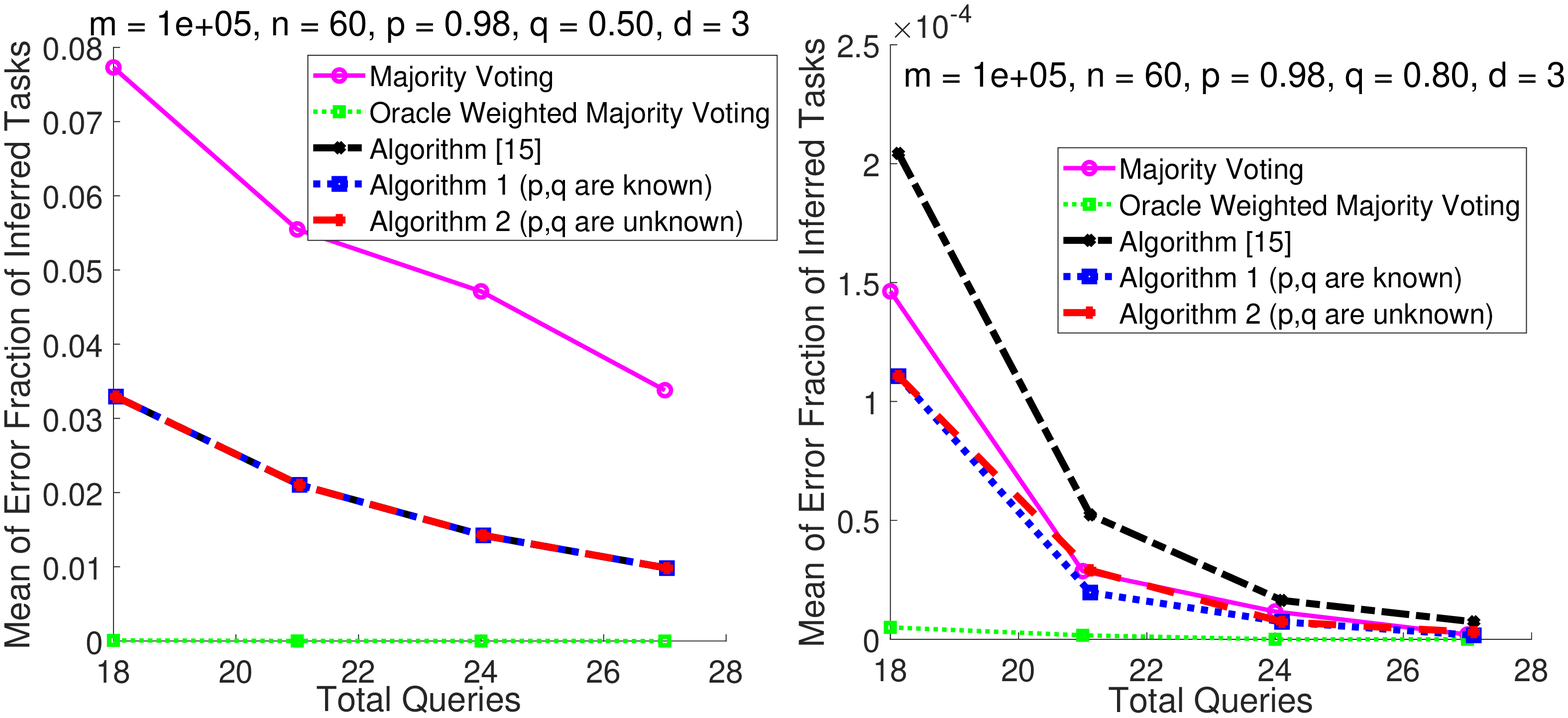}
	\caption{Comparisons of label recovery accuracy for five different algorithms.}
	\label{fig:comp1}
	\vspace{-0.3cm}
\end{figure}

We provide simulation results to show that the proposed algorithms outperform  other baselines in diverse parameter regimes. 
In Fig.~\ref{fig:comp1}, we compare our algorithm ({Alg.1 and 2}) with majority voting, oracle weighted majority voting, and {Alg.~\cite{shah2018reducing}} in terms of 
the error fraction in inferred tasks over the number of queries per task when $d=3$. The result is averaged over 30 times Monte Carlo simulations.
When $q=1/2$ (left figure), {Alg. 1} becomes the same as {Alg.~\cite{shah2018reducing}} and these algorithms outperform the majority voting. We can observe that {Alg. 2}, which uses the estimates $(\hat{p},\hat{q})$, achieves as good performance as that of {Alg. 1}. 
When $q>1/2$ (right figure), our algorithms show the best performance. 

\begin{figure}
	\centering
	\includegraphics[width=7cm]{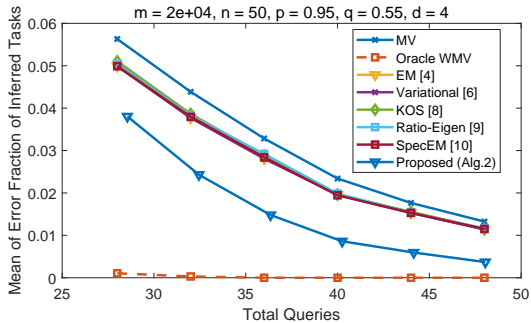}
	\caption{Comparison of the proposed algorithm with the state-of-the-art algorithms designed for the single-coin Dawid-Skene model.}
	\label{fig:comp2}
	\vspace{-0.4cm}
\end{figure}

In Fig.~\ref{fig:comp2}, the performance of the proposed algorithm is compared with that of the state-of-the-art algorithms developed for the single-coin Dawid-Skene model, which assumes that the worker reliability does not change depending on the task type. The state-of-the-art algorithms perform worse than the proposed algorithm when the data is collected assuming the worker-type specialization model, which may reflect more realistic scenarios in diverse crowdsourcing applications. 

\section{Conclusions}\label{sec6}
We considered crowdsourced labeling under a $d$-type specialization model with general reliability parameters $p> q\geq 1/2$. When $(p,q)$ values are known but not the types of tasks/workers, our proposed algorithm (Alg. 1) recovers binary tasks up to any given accuracy $(1-\alpha_c)\in(0,1)$ with the number of queries per task scales as $\Theta(d\ln\frac{d}{\alpha_c})$ when $q=1/2$ and as $\Theta(\ln\frac{d}{\alpha_c})$ when $q>1/2$. 
We also proposed an algorithm (Alg. 2) that does not require any information about reliability parameters nor the task/worker types, and empirically showed that this algorithm achieves as good performance as the algorithm with the known reliability parameters $(p,q)$. 
\appendices

\bibliographystyle{IEEEtran}
\bibliography{IEEEabrv,21ISIT_crowd_cluster}

\end{document}